# Crystal Electrostatic Energy

Alexander Ivanchin

> It has been shown that to calculate the parameters of the electrostatic field of the ion crystal lattice it sufficient to take into account ions located at a distance of 1-2 lattice spacings. More distant ions make insignificant contribution. As a result, the electrostatic energy of the ion lattice in the alkaline halide crystal produced by both positive and negative ions is in good agreement with experiment when the melting temperature and the shear modulus are calculated. For fcc and bcc metals the ion lattice electrostatic energy is not sufficient to obtain the observed values of these parameters. It is possible to resolve the contradiction if one assumes that the electron density is strongly localized and has a crystal structure described by the lattice δ- function. As a result, positive charges alternate with negative ones as in the alkaline halide crystal. Such δ-like localization of the electron density is known as "a model of nearly free electrons".

**Introduction**

There are some problems in the calculation of the ion lattice electrostatic energy. Direct summation of ion potentials leads to divergent or conditionally convergent series. The above series have no sum and it is incorrect to perform any mathematic operations on them. Some examples are shown in [1]. In spite of the fact, Evald carried out calculation. For series that do not converge absolutely the Fourier transformation is incorrect. It is not clear if it has anything to do with the reality of Evald's calculation. That is what J. Ziman [2] states directly. For correct calculation it is necessary to clear up the problem.

**1. Two-dimension case**

Suppose positively charged straight lines are located in the plane $xz$ in parallel to the axis $y$. The intensity of the straight line passing through point $nl$ on the abscissa is (here $w = x + iy$) [3].

$$E_n = \frac{\rho}{2\pi\varepsilon} \frac{1}{\overline{w} - nl} \tag{1.1}$$

where $\rho$ is the linear charge density, $\varepsilon$ is the electric constant. The bar stands for a conjugate complex number. For a finite number of straight lines summing (1.1) we obtain

$$E(N) = \frac{\rho}{2\pi\varepsilon l} \sum_{n=-N}^{+N} \frac{1}{(\overline{w}/l) - n} \tag{1.2}$$

At $N \to \infty$ the series in (1.2) diverges. Let us transform (1.2) combining the components with the numbers $n$ and $-n$

$$E(N) = \frac{\rho}{2\pi\varepsilon l} \left[ \frac{l}{\overline{w}} + \frac{\overline{w}}{l} \sum_{n=1}^{+N} \frac{2}{(\overline{w}/l)^2 - n^2} \right] \tag{1.3}$$

At $N \to \infty$ the series converges absolutely [4]

$$E = \lim_{N \to \infty} E(N) = \frac{\rho}{2\pi\varepsilon l} \left[ \frac{l}{\overline{w}} + \frac{\overline{w}}{l} \sum_{n=1}^{+\infty} \frac{2}{(\overline{w}/l)^2 - n^2} \right] = \frac{\rho}{2\varepsilon l} \mathrm{ctg}\left(\pi \frac{\overline{w}}{l}\right) \tag{1.4}$$

In the real form (1.4) is written as

$$E_x = \frac{\rho}{4\pi\varepsilon l} \frac{\sin(2\pi x')}{\sin^2(\pi x')\cosh^2(\pi y') + \cos^2(\pi x')\sinh^2(\pi y')} \approx$$

$$\approx \frac{\rho}{2\pi\varepsilon l} \exp(-2\pi y')\sin(2\pi x')$$

$$E_y = \frac{\rho}{4\pi\varepsilon} \frac{\sinh(2\pi y')}{\sin^2(\pi x')\cosh^2(\pi y') + \cos^2(\pi x')\sinh^2(\pi y')} \approx$$

$$\approx \frac{\rho}{2\pi\varepsilon l}[1 + \exp(-2\pi y')\cos(2\pi x')]$$

(1.5)

Here

$$x' = x/l, \quad y' = y/l, \quad z' = z/l. \tag{1.6}$$

The equation (1.5) gives an approximate value of intensity accurate to the 2$^{nd}$ order. The amplitude of fluctuation of the electrostatic field is

$$A = \frac{\rho}{4\pi\varepsilon l}\exp(-2\pi y') \tag{1.7}$$

One can see that at a distance of only one lattice spacing $(y' = 1)$ the electrostatic field is practically similar to the field of a uniformly charged plane. Then a question arises about the case of a three-dimension crystal. Further it is shown that the same is true for the latter case.

**2. Point charges**

Let point positive charges be located at points $nL$ on the ordinate and $e$ is the electron charge. Suppose $|y| \leq L/2$. The potential at the point $(x, nL - y, z)$ is written as

$$\varphi = \frac{e}{4\pi\varepsilon} \sum_{n=-N}^{N} \frac{1}{\sqrt{x^2 + (nL - y)^2 + z^2}} \tag{2.1}$$

At $N \to \infty$ the sum in (2.1) diverges.

Let us consider a uniformly charged straight-line segment $L[n - 1/2, \; n + 1/2]$ located on the ordinate of the length $L$ with the linear charge density $e/L$ and the center at point $nL$. Its potential is written as

$$\varphi_n = \frac{e}{4\pi\varepsilon L} \int_{-L/2}^{L/2} \frac{d\xi}{\sqrt{(\xi + nL - y)^2 + r^2}} = \frac{e}{4\pi\varepsilon L} \int_{-v_n}^{v_n} \frac{dt}{\sqrt{1 - 2c_n t + t^2}} =$$

$$= \frac{e}{4\pi\varepsilon L} \ln \frac{\sqrt{v_n^2 - 2v_n c_n + 1} + v_n - c_n}{\sqrt{v_n^2 + 2v_n c_n + 1} - v_n - c_n}$$

(2.2)

Here

$$r = \sqrt{x^2 + z^2}$$

$$R_n = \sqrt{(y-nL)^2 + r^2}$$

$$c_n = \frac{y-nL}{R_n} \qquad (2.4)$$

$$v_n = \frac{L}{2R_n}$$

$$t = \frac{\xi}{R_n}$$

Expanding the integrand into a series in the Legendre polynomials at $v_n < 1$ we obtain [4]

$$\varphi_n = \frac{e}{4\pi\varepsilon L} \int_{-v_n}^{v_n} \frac{dt}{\sqrt{1-2c_n t + t^2}} = \frac{e}{4\pi\varepsilon L} \sum_{k=0}^{\infty} \int_{-v_n}^{v_{n_{mn}}} P_k(c_n) t^k dt =$$

$$= \frac{e}{2\pi\varepsilon L} \sum_{k=0}^{\infty} \frac{P_{2k}(c_n)}{(2k+1)} v_n^{2k+1} = \frac{e}{4\pi\varepsilon R_n} + \frac{e}{2\pi\varepsilon L} \sigma_n \qquad (2.5)$$

Here

$$\sigma_n = \sum_{k=1}^{\infty} \frac{P_{2k}(c_n)}{2k+1} v_n^{2k+1} \qquad (2.6)$$

$P_k(c_n)$ is the Legendre polynomial of the $k$ th order. From (2.2) and (2.5) we get

$$\ln \frac{\sqrt{v_n^2 - 2v_n c_n + 1} + v_n - c_n}{\sqrt{v_n^2 + 2v_n c_n + 1} - v_n - c_n} - v_n = 2\sigma_n \qquad (2.7)$$

The equation (2.7) shows that the point charge potential and that of the segment differ by $2\sigma_n$. The value $\sigma_n \sim v_n^3/3$ at small values of $v_n$. The calculated values of $\sigma_n$ depending on the distance from the plane with point charges are shown in Table 1.

Table 1

| $R_n$ | L | 2L | 3L | 4L | 5L |
|---|---|---|---|---|---|
| $v_n$ | 1/2 | 1/4 | 1/6 | 1/8 | 1/10 |
| $\sigma_n$ | 0.046 | 0.0052 | 0.0015 | 0.00065 | 0.00033 |

At a distance of only one lattice spacing the field of the charged segment is not very different from the field of the point charge. Summing (2.7) over the whole chain we get

$$\sum_{n=-N}^{N}\left[\ln\frac{\sqrt{v_n^2-2v_nc_n+1}+v_n-c_n}{\sqrt{v_n^2+2v_nc_n+1}-v_n-c_n}-v_n\right]=2\sum_{n=-N}^{N}\sigma_n \qquad (2.8)$$

The first term in the left part of (2.8) changes over to the potential of the infinite straight line at $N\to\infty$. The problem on point charges in the plane has been reduced to that on charged straight lines.

Let us place negative ions at points $(n+1/2)L$ in the middle between the positive ions. For them the obtained relations will be the same. One should only replace $y\to y-L/2$ and change the charge sign of ions. As a result, the field, as a matter of fact, disappears at a distance of only one lattice spacing $L$ from the line with the chain of sign-variable charges.

The planes (100) in the alkaline halide crystal consist of the above sign-variable chains. The lattice constant $a$ in the alkaline halide crystal is the nearest distance between two unlike ions. The most close-packed direction of the same ions in the alkaline halide crystal is [110] with distances between the ions $L=\sqrt{2}a$. The chains are located in the planes (100). Let us calculate the potential drop $\Delta\varphi(x)$ between point $O$ located at the origin of the coordinates and an arbitrary point $M$ with coordinates (0, $y$ ,0) located in [110]. The latter direction is shown in green in Fig.1.

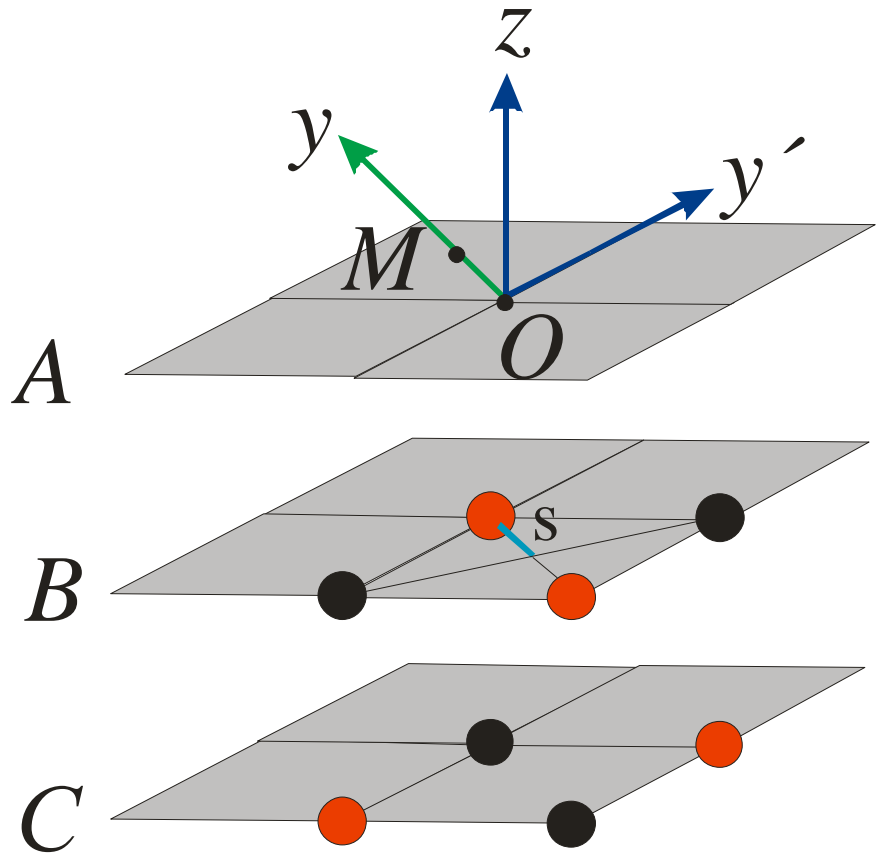

Fig.1. Planes (100) in the alkaline halide crystal. Only a few ions are shown for simplicity of the drawing.

The chain [110] nearest to point $O$ is located in the plane $B$ at a distance a and consists, for example, of positive ions. The positive ions are marked in red and the negative ones in black. The nearest chains of the second order are also in the plane $B$ at a distance of $a\sqrt{3}/2 < L$ from point $O$ and consist of negative ions. The next two nearest chains [110] in the same plane are located at a distance of $a\sqrt{3} > L$, therefore it can be neglected. The plane $C$ is at a distance $2a > L$, and it can also be neglected. The potential drop is written as

$$\varphi = \frac{e}{4\pi\varepsilon} \sum_j \sum_{n=-N}^{N} \left( \frac{(-1)^j}{\sqrt{(y-nL+s)^2 + r_j^2}} - \frac{1}{\sqrt{(y_0-nL+s)^2 + r_j^2}} \right) = $$

$$= \frac{e}{4\pi\varepsilon a} \sum_j \sum_{n=-N}^{N} \left( \frac{(-1)^j}{\sqrt{(\chi-n\lambda+\gamma)^2 + \rho_j^2}} - \frac{1}{\sqrt{(\chi_0-n\lambda+\gamma)^2 + \rho_j^2}} \right) \quad (2.9)$$

The values in {2.9) are as follows

$$\chi_0 = 0, \rho_0 = 1, \quad \rho_1 = \rho_{-1} = \sqrt{\frac{3}{2}}, \quad \lambda = \frac{L}{a} = \sqrt{2}, \quad \gamma = \frac{s}{a} = \frac{\sqrt{2}}{2}$$

$\chi$ covers the values from 0 to $\sqrt{2}$, $s = 0$ for positive ions and $s = a\sqrt{2}/2$ for negative ions, $s$ takes into account the shift of unlike chains relative to each other along the axis $y$ and is marked in blue in Fig.1, $r_j$ takes into account the sign of the charges in the chain. If all the chains consist of charges of the same sign, then the multiplier $(-1)^j$ should be replaced by 1. Summation over $n$ is performed over the ions from chain $j$, while summation over $j$ is performed over the chains. The maximum value of $\varphi(y)$ on the axis $Oy$ is achieved at point $M$ at a distance of $\chi = \sqrt{2}/2$ from point $O$ and is equal to

$$\varphi_M = \max \sum_j \sum_{n=-N}^{N} \left( \frac{(-1)^j}{\sqrt{(\chi-n\lambda)^2 + \rho_j^2}} - \frac{1}{\sqrt{(\chi_0-n\lambda)^2 + \rho_j^2}} \right) \approx 0.066 \quad (2.10)$$

For accuracy exceeding 1 % it is sufficient to take $N = 10$. Calculation shows that the contribution of the nearest chain located, at a distance $a$ is 80 %, whereas that of the nearest chains of the second order ~ 19% and that of the rest less than 1 %. Therefore, to calculate the potential to an accuracy up to 99 % in the alkaline halide crystal it is enough to take into account chains of the first and second order located in the nearest plane.

The height of the energy barrier, which an ion has to overcome when passing from one equilibrium state into another will be

$$w = e\varphi_M \tag{2.11}$$

The formula (2.11) is related to a case when the plane $A$ is a free surface. If the plane $A$ is situated deep inside a crystal, then the energy value in (2.11) should be doubled to take into account the influence of the plane $D$ shown in Figure 2.

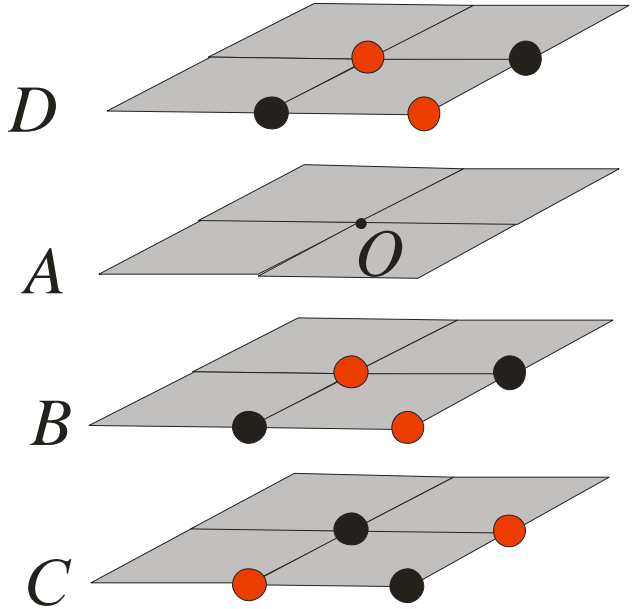

Figure 2

As a result, we get

$$w = 0.066 \frac{B}{a} \tag{2.12}$$

Here

$$B = \frac{e^2}{4\pi\varepsilon} \approx 14{,}35 \cdot 10^{-10} \, eV \cdot m \tag{2.13}$$

Above the Debye temperature an ion, on the average, has an energy equal to $3kT$. If we add the heat of melting falling at one ion $q$ to the energy $w$ in (2.12), then we derive an average energy per one ion at which transition from the solid state into the liquid state occurs, i.e. we obtain theoretical evaluation of the melting temperature $T_{th}$ using the formula

$$T_{th} = \frac{q+w}{3k} \tag{2.14}$$

It should be noted that when calculating the height of the energy barrier (2.12) the ions were considered to be point charges. As a matter of fact, due to thermal fluctuations the effective ion charge will be a little delocalized near the lattice point, which decreases the energy height.

Evaluation of the melting temperature without taking into account delocalization will yield overestimated values. In Table 2 the experimentally measured values of the alkaline halide crystal melting temperature $T_e$ are compared with the calculation data $T_{th}$. The evaluation yields a reasonable although somewhat overestimated value of the melting temperature.

Table 1. The alkaline halide crystal melting temperatures: calculation and experiment

|  | Lattice constant $a$ in $nm$ | Heat of melting in kJ/mol | Heat of melting $q$ in $eV$ per ion | $w$ $eV$ | $q+w$ $eV$ | $T_{th}$ $K$ | $T_e$ $K$ |
|---|---|---|---|---|---|---|---|
| NaCl | 0.5640 | 28.8 | 0.15 | 0.17 | 0.32 | 1185 | 1074 |
| LiF | 0.4026 | 26.4 | 0.138 | 0.242 | 0.38 | 1472 | 1143 |
| KCl | 0.6292 | 25.5 | 0.133 | 0.155 | 0,288 | 1116 | 1049 |
| KBr | 0.6598 | 24.8 | 0.129 | 0.147 | 0.276 | 1069 | 1001 |
| KF | 0.46344 | 34.25 | 0.178 | 0.210 | 0.388 | 1500 | 1269 |
| LiCl | 0.51398 | 13.,4 | 0.0678 | 0.190 | 0.2578 | 999 | 887 |

## 3. Face-centered cubic lattice

In fcc the most close-packed directions are [110] (Fig. 3)

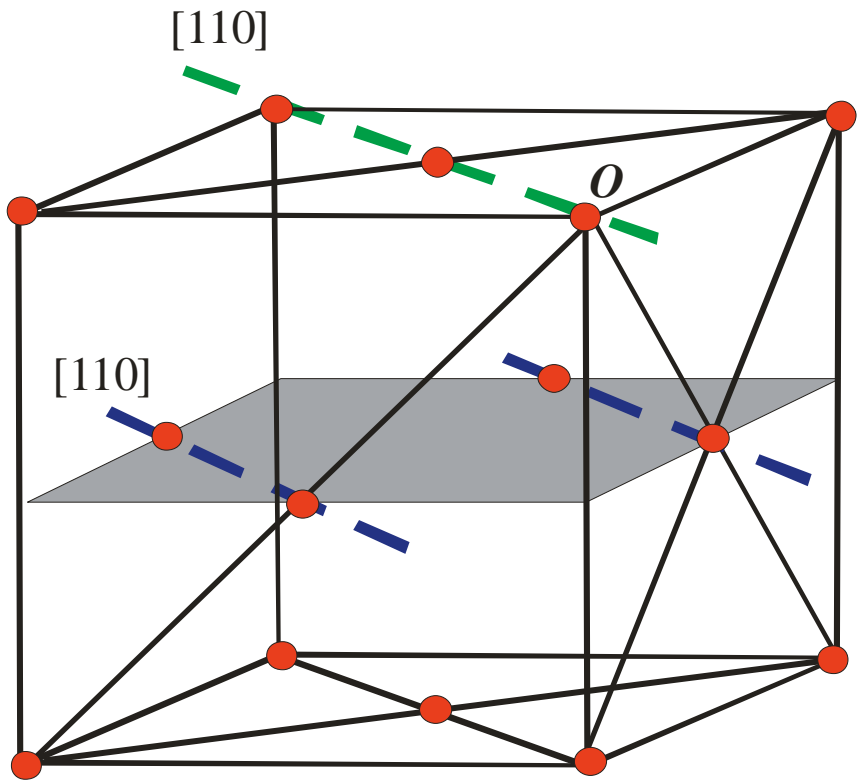

Figure 3

The distance between the ions in that direction is $L = a\sqrt{2}/2$. The lattice constant $a$ is the edge length of a face-centered cube. The nearest chains in the direction [110] from point $O$ lie in the plane (100), they are designated by a blue broken line and located at a distance $b_1 = a\sqrt{6}/4$. The relation $b_1/L = \sqrt{3}/2 \approx 0{,}87$. The nearest chains of the second order are at a distance $b_2 = a\sqrt{2}/2$, for them the relation $b_2/L = 1$ and its influence can be neglected. For fcc using (2.10) we derive

$$\varphi_M = 2\max \sum_{n=-N}^{N}\left(\frac{1}{\sqrt{(\chi - n\lambda + \gamma)^2 + \rho^2}} - \frac{1}{\sqrt{(n\lambda + \gamma)^2 + \rho^2}}\right) \approx 0.0046 \qquad (3.1)$$

Here $\rho = \sqrt{6}/4$, $\lambda = \sqrt{2}/2$, $\gamma = \sqrt{2}/4$. Unlike the alkaline halide crystal, only two nearest similar chains should be taken into account on the free surface, therefore, summation over $j$ is replaced by coefficient 2. The height of the energy barrier is

$$w = 0.0046 \frac{B}{a} \qquad (3.2)$$

For Al the lattice constant $a = 0{,}404\ nm$, from (3.2) we obtain $w \approx 0{,}016\ eV$. At room temperature, on the average, the atom has $3kT \sim 0{,}08\ eV$. Hence, the height of the potential barrier produced by the fcc ion lattice is small and cannot ensure its existence in the solid state.

## 4. Body-Centered Cubic Lattice

For bcc the most close – packed direction is [111] (Fig. 4)

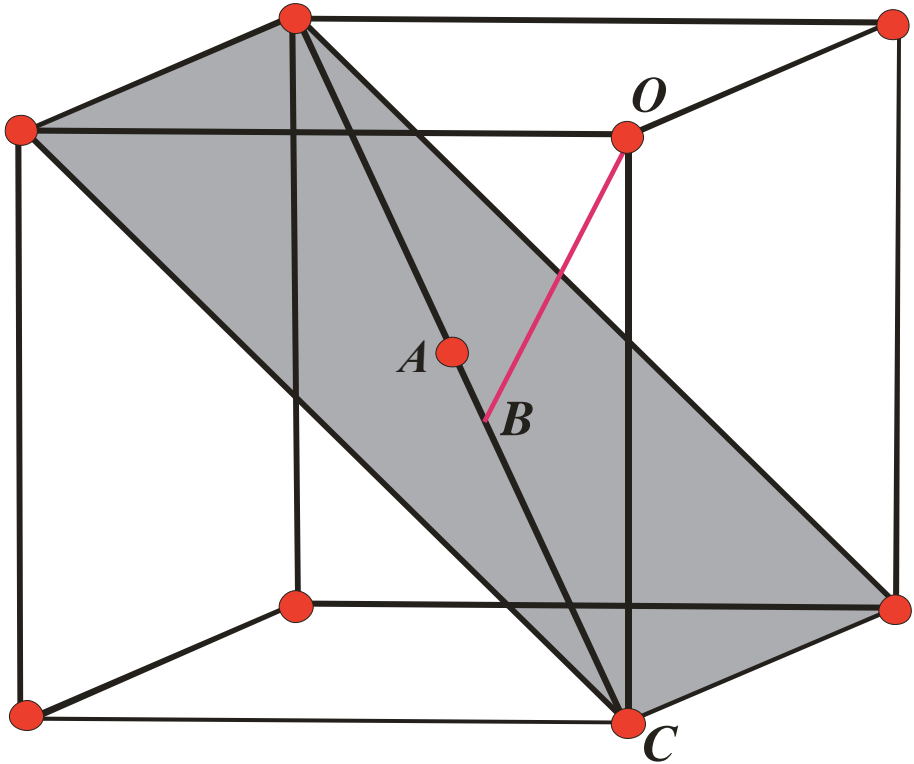

Figure 4

The distance between the ions $L = a\sqrt{3}/2$. The distance between the nearest directions [111] in bcc is the segment OB of the length $a\sqrt{2/3}$. The relative distance between the chains $a\sqrt{2/3}/L = 2\sqrt{2}/3 \approx 0{,}94$ is close to unity. Therefore, it is enough to take into account only one chain. Using (2.10) we obtain (here $\gamma = \sqrt{2}/4$)

$$\varphi_M = 0.012 \qquad (4.1)$$

The height of the energy barrier on the free surface is
$$w = 0.012 \frac{B}{a} \tag{4.2}$$

For tungsten $a = 0.3159\ nm$, then
$$w \approx 0{,}05\ eV \tag{4.3}$$

For tungsten the height of the energy barrier produced by the ion lattice is also small and cannot ensure the solid state at room temperature.

## 5. Shear modulus

Elastic deformation is determined in the following way
$$\eta_{yz} = \frac{1}{2}\left(\frac{\partial U_z}{\partial y} + \frac{\partial U_y}{\partial z}\right) \tag{5.1}$$

Here **U** is the vector of elastic displacement. For displacement considered here only component $U_y$ is not zero and depends on the coordinate z. Then (5.1) looks like
$$\eta_{yz} = \frac{1}{2}\frac{dU_y}{dz} \tag{5.2}$$

This is the only component of the deformation tensor which is not equal to zero, therefore, below we omit its indices. $U_y$ is the ion displacement and in (2.9) is designated by $y$. A transition to the discrete medium is implemented by the replacement of $dU_y$ in (5.2) by $y$ and $dz$ by the distance between the nearest close-packed chains $b_1$. Then the equation (5.2) is written as
$$\eta = \frac{1}{2}\frac{y}{b_1} = \frac{1}{2}\frac{\chi}{\mu} \tag{5.3}$$

where $\mu = b_1/a$ is the dimensionless interplane distance, $\mu = \sqrt{3}/2$ for planes (110) in bcc, $\mu = 1/2$ for planes (100) in fcc and $\mu = 1$ for the planes (100) in the alkaline halide crystal.
From (5.2) and (5.3) the rules of differentiation follow
$$\frac{\partial}{\partial \eta} = \frac{\partial}{\partial \chi}\frac{\partial \chi}{\partial \eta} = 2\mu\frac{\partial}{\partial \chi}$$
$$\frac{\partial^2}{\partial \eta^2} = 4\mu^2 \frac{\partial^2}{\partial \chi^2} \tag{5.4}$$

When calculating the shear modulus one should also take into account 1-3 nearest chains, as when evaluating the melting temperature. To determine the elastic modulus, the energy density is used rather than the energy per atom. Therefore, the energy per atom should be divided by the volume per one ion $v$. For the alkaline halide crystal $v = a^3$, for fcc $v = a^3/4$ and for bcc $v = a^3/2$. As a result, we obtain
$$\tilde{w} = \frac{2e\varphi}{v} \tag{5.5}$$

Since there are two nearest planes inside the crystal, there appears coefficient 2 in (5.5). Using (5.1)-(5.5) the shear modulus is written as

$$G_{2323} = \left.\frac{\partial^2 \tilde{w}}{\partial \eta^2}\right|_{\eta=0} = \left.\frac{\partial^2 (2e\varphi)}{\nu \partial \eta^2}\right|_{\eta=0} = \frac{2e^2 \mu^2}{\pi \varepsilon \nu a} \frac{\partial^2}{\partial \chi^2} \sum_j \sum_{n=-N}^{N} \left.\frac{(-1)^j}{\sqrt{(\chi - n\lambda + \gamma)^2 + \rho_j^2}}\right|_{\chi=0} = $$

$$= \frac{2e^2 \mu^2}{\pi \varepsilon \nu a} S \qquad (5.6)$$

Here it is designated

$$S = \sum_j \sum_{n=-N}^{N} (-1)^j \frac{2(n\lambda - \gamma)^2 - \rho_j^2}{\left[(n\lambda - \gamma)^2 + \rho_j^2\right]^{5/2}} \qquad (5.7)$$

As one component $G_{2323}$ of the elastic modulus tensor is evaluated, its indices are omitted. For the alkaline halide crystal the calculation yields
$$S \approx 1{,}05 \qquad (5.8)$$

Here $\lambda = \sqrt{2}$, $\gamma = 0$, $\rho = 1$ for the 1st order chain and $\gamma = \sqrt{2}/2$, $\rho = \sqrt{3/2}$ for the chains of the second-order closeness and for the third-order chains in the nearest plane $\gamma = 0$, $\rho = \sqrt{3}$.
The other chains may not be taken into account. The contribution of the electrostatic component of the ion lattice in the alkaline halide crystal into the shear modulus is described by

$$G = \frac{2e^2}{\pi \varepsilon a^4} S \approx 2{,}16 \frac{e^2}{\pi \varepsilon a^4} \qquad (5.9)$$

Here we have obtained the value of the elastics modulus for the plane (100) in the direction [110].
   In order to pass to the coordinate system determined by the elementary cube it is necessary to turn the coordinates by an angle of $\pi/4$ round the axis $z$ (Fig.1), then the axis $y'$ will take the position of the axis $y$. $G$ turned out to be of the same value in these coordinate systems. The calculated and measured values of $G$ are given in Table 2.

Table 2

| Name | Lattice constant $a$ in nm | $G$ Theory $\cdot 10^{10}\ N/m^2$ | $G$ Experiment $\cdot 10^{10}\ N/m^2$ |
|------|---------------------------|-----------------------------------|---------------------------------------|
| NaCl | 0.5640 | 1.97 | 1.59 |
| LiF  | 0.4026 | 7.59 | 4.09 |
| KCl  | 0.6292 | 1.27 | 1.08 |
| KBr  | 0.6598 | 1.05 | 0.90 |

As is has been noted above, the theoretical value of $G$ is higher than in the experimental data.
   For fcc it is necessary to take into account only four nearest chains, their contribution being the same. The shear modulus will be written as

$$G = \frac{e^2}{4\pi \varepsilon a^4} S = \frac{B}{a^4} S \qquad (5.10)$$

$$S = 4\sum_{n=-N}^{N} \frac{2(n\lambda - \gamma)^2 - \rho^2}{\left[(n\lambda - \gamma)^2 + \rho^2\right]^{5/2}} \approx 3.8$$

Since four nearest chains make the same contribution, then summation over $j$ is replaced by multiplier 4. For copper the lattice constant $a = 0,36$ $nm$, then using (5.10) we obtain
$G = 3,4 \cdot 10^9$ $N/m^2$

and the measured value is
$G = 4,2 \cdot 10^{10}$ $N/m^2$

That is, the shear modulus in fcc metals is not determined by the electrostatic energy of the ion lattice.

For bcc metals the distance between the ions in the most close-packed directions [111] is $L = a\sqrt{3}/2$, the nearest distance between the chains [111] is $\rho a = a\sqrt{2/3}$, then $\rho a / L = 2\sqrt{2}/3 \approx 0,94$. The above value is closer to 1 than for fcc. $\gamma = 1/2\sqrt{3}$, $v = a^3/2$, $\rho = \sqrt{2/3}$, $\lambda = \sqrt{3}/2$. Then

$$G = \frac{3e^2}{\pi \varepsilon a^4} = \frac{3e^2}{\pi \varepsilon a^4} S$$

$$S = \sum_{n=-N}^{N} \frac{2(n\lambda - \gamma)^2 - \rho^2}{\left[(nL - \gamma)^2 + \rho^2\right]^{5/2}}$$

As is the case with fcc, summation over $j$ is replaced by multiplier 2, because there are two nearest chains. As a result of the calculation, we obtain $S \approx 0,36$, and the shear modulus is by an order of magnitude less than the measurement data. The ion lattice in bcc yields an insignificant contribution into the shear modulus.

In addition, projection of point O on [111], and namely, point B, does not lie in the middle of the segment AC (Fig.4). Therefore, the first derivative in the direction of [111] from the energy does not reduce to zero. It can be verified by calculation. It turns out that the directions [111] are not in equilibrium. If the electron density is spread over the elementary cell ("electron jelly"), then there are no forces keeping the bcc lattice in equilibrium.

**6. Discussion**

The main results of the calculations presented here are as follows
1. The electrostatic field of ions located regularly on the plane coincides, in fact, with that of the uniformly charged plane at a distance of only one lattice spacing between the ions.
2. To calculate the parameters of the crystal electrostatic field it is sufficient to take into account only the nearest ions.
3. The electrostatic energy of the ion lattice in fcc and bcc metals is too small to ensure the existence of the crystals in the solid state at room temperature.
4. The electrostatic energy of the ion lattice is too small to create the elastic shear modulus in fcc and bcc metals.
5. The electrostatic energy of the ion lattice of a bcc metal cannot ensure equilibrium in the form of a cube.
6. The alkaline halide crystal is free from such contradictions.

The only way to get rid of the above contradictions is to reject the idea of the electron density distributed over the elementary cell volume ("electron jelly"). Suppose electrons like ions form a crystal structure. One of the variants is that in fcc metals the electron sublattice replicates the ion structure and is shifted relative to it in the direction [110] at a distance

$$a\sqrt{2}/2. \tag{6.1}$$

Then the ion sublattice in fcc together with the electron sublattice form a structure similar to that of the alkaline halide crystal. In the case of bcc, it is not so easy to make assumptions on the structure.

The hypothesis of the $\delta$-like localization of the electron density is in good agreement with the theory of the metal electron structure and its main idea of "nearly free electrons". According to the theory, the density of the electron probability in metals is the plane wave

$$C\exp(i\mathbf{g}\cdot\mathbf{x}) \tag{6.2}$$

Here

$$\mathbf{g} = \frac{2\pi}{a}\{n, \quad m, \quad k\} \tag{6.3}$$

where the vector of the reciprocal lattice - $n$, $m$, $k$ is natural numbers that achieve large values. That $n$, $m$, $k$ are natural numbers is the consequence of periodicity of the crystal lattice. The electron density in metals is the superposition of plane waves (6.2)

$$C\sum_{g}\exp(i\mathbf{g}\cdot\mathbf{x}) \tag{6.4}$$

Summation is performed over all the vectors of the reciprocal lattice. In mathematics there is the well-known Poisson formula

$$\sum_{k=-\infty}^{+\infty}\exp(2k\pi ix) = \sum_{k=-\infty}^{+\infty}\delta(x-k) \tag{6.5}$$

In (6.5) unlike (6.4) summation is performed infinitely. However, since the series

$$\sum_{k=-\infty}^{+\infty}\exp(2k\pi ix) \tag{6.6}$$

converges, at least in the sense of generalized functions [5], it is not essential.

Thus, the electron density distribution in metals is the lattice $\delta$-function. The name "nearly free electrons" is, probably, humorous. In order to avoid the $\delta$-like localization of the electron density, the vector $g$ in (6.2) should pass over a continuous rather than discrete series of values.

Localization of the electron density does not contradict the principle of uncertainty, which is the result of interference, and the quantum mechanics does not forbid it.

The electron system can form several variants of the crystal lattice that have various energetics, and it does not necessarily replicate the ion lattice structure. With increasing temperature the electron crystal structure may lose stability and change over to a more suitable variant from the point of view of energy. It leads to a change in the ion sublattice, i.e. to phase transition. The structure of the electron sublattice can be influenced by various factors, for example, the ion charge, the ion mass, etc.

The hypothesis of the electron lattice eliminates the above problems. The constant of the alkaline halide structure, which is obtained from the fcc lattice by adding the electron sublattice, is $a_0 = a/2$. The height of the potential barrier will be calculated using (2.12) and the elastic modulus by means of (5.9). They are more than sufficient to ensure a high temperature of melting metals and the measured elastic modulus.